\documentclass[preprint]{JHEP3} 
\usepackage{graphicx} 
\usepackage{epsfig}
\usepackage{amsmath} 
\usepackage{bbm} 
\usepackage{cite}
\newcommand{\der}[2]{\ensuremath{\frac{\partial #1}{\partial #2}}}

\title{Topology with Dynamical Overlap Fermions} 
\author{ G. I. Egri$^{a,b}$, Z. Fodor$^{a,b}$, S. D. Katz$^{a}$ and
K.K. Szab\'o$^{b}$\\ \it $^a$Institute for Theoretical Physics, E\"otv\"os
University, Budapest, Hungary\\ \it $^b$Department of Physics, University of
Wuppertal, Germany}

\date{\today} 
\abstract{ We perform dynamical QCD simulations with
$n_f=2$ overlap fermions by hybrid Monte-Carlo method on $6^4$ to $8^3\times
16$ lattices.  We study the problem of topological sector changing. A new
method is proposed which works without topological sector
changes.  We use
this new method to determine the topological susceptibility at various quark
masses.}

\preprint{hep-lat/0510117}


\begin{document}

\section{Introduction}

The overlap operator \cite{Neuberger:1997fp,Neuberger:1998wv} , gives a
theoretically sound solution of the chirality problem on the lattice.  It
satisfies the Ginsparg-Wilson relation
\cite{Ginsparg:1981bj,Hasenfratz:1997ft}, which ensures the exact chiral
symmetry at finite lattice spacing \cite{Luscher:1998pq}, moreover the
difference in the number of left and right handed zero modes can be taken as a
definition of the topological charge ($Q$) which gives the correct result in
the continuum limit \cite{Hasenfratz:1998ri}.

However, the numerical implementation of dynamical overlap fermions is still a
great challenge today (for early studies with dynamical overlap fermions we
refer to \cite{Narayanan:1995sv,Neuberger:1999re,Bode:1999dd}).  The presence
of nested conjugate gradients for the inversion of the Dirac operator makes
the simulations considerably slower than simulations with Wilson
fermions. Furthermore one has to face the non-continuity of the fermion
determinant at the boundary of topological sectors. This additional difficulty
can be treated exactly in frame of the Hybrid Monte Carlo (HMC) algorithm by
modifying the molecular dynamics trajectory at the boundary
\cite{Fodor:2003bh}. Clearly the crossing rate between different topological
sectors is heavily affected by this modification. Inappropriate treatment
might confine the system into a certain topological sector which yields an
unacceptably large autocorrelation time for $Q$ in the simulation.

A few exploratory studies are already available in QCD with dynamical overlap
fer\-mi\-ons
\cite{Arnold:2003sx,Fodor:2003bh,Cundy:2004pz,Cundy:2005pi,DeGrand:2004nq,DeGrand:2005vb}.
All handle the modification of the trajectory at the boundary in a similar
style. The original proposal of \cite{Fodor:2003bh} is modified in
\cite{Cundy:2005pi} in such a way that the acceptance rate is increased.  It
is shown in \cite{DeGrand:2004nq}, that the introduction of several
pseudofermion fields which approximate the fermion determinant, can enhance
the crossing rate.  The relation between the pseudofermionic (over)estimation
and rare topological sector changes\footnote{In the staggered formulation
there was already a concern that the pseudofermion estimator obstruct the
change of topological charge \cite{Dilger:1994ma}.}  was pointed out in
\cite{DeGrand:2005vb}.

In this paper we study the problem of changing topological sectors in the case
of the overlap operator with $n_f=2$ fermions in dynamical HMC
simulations. Sec. 2 will give a short introduction to the sector changing
problem for overlap fermions (by summarizing and extending the work of
\cite{DeGrand:2005vb}), and answers the question why the present treatment is
unlikely to change $Q$. In Sec. 3 a new measurement method of expectation
values is proposed, which circumvents the crossing problem entirely by making
simulations constrained to fixed topological sectors. In Sec. 4 we present
numerical results using the new measurement method. In Sec. 5 conclusions are
given.

\section{Topological sector changing problem}

\label{problem}

After introducing pseudofermion fields \cite{Duane:1987de}, our partition
function reads:
\begin{equation}
Z= \int [dU] e^{-S_g} \det H^2 = \int [dU] [d \phi ^ \dagger] [d \phi] e^{-S_g
  - \phi ^ \dagger H^{-2} \phi},
\end{equation}
where \(H\) is the hermitian overlap operator:
\begin{equation}
H = (1-\frac{m}{2 m_0}) H_0 + \gamma_5 m,
\end{equation}
with \(H_0\) being the massless hermitian overlap operator:
\begin{equation}
H_{0}=m_0 [\gamma_5 + \textrm{sgn}(H_W)].
\end{equation}
Here $H_W$ is the standard Wilson operator with negative mass $-m_0$.  The sgn
function in \(H_0\) causes a Dirac-\(\delta\) type singularity in the equation
of motion of the momenta of the link variables. The \(\delta\)-function gives
a contribution whenever an eigenvalue \(\lambda\) of \(H_{W}\) changes
sign. This subspace of the configuration space coincide with topological
sector boundaries. The reason for this is that in the case of the overlap
operator the topological charge is:
\begin{equation}
Q = \frac{1}{2 m_0} \textrm{Tr}(H_0)=\frac{1}{2} \sum_i \textrm{sgn}(
\lambda_i) 
\end{equation}

This means, that there are potential walls, non-differentiable steps in the
action at the topological sector boundaries. The reflection-refraction method
suggested in \cite{Fodor:2003bh} handles these potential walls
correctly. Let's denote the momenta by \(p\) and the normal vector of the
topological sector boundary by \(n\). According to this method one has to
modify the momenta, when arriving at a potential wall:

\begin{equation}
\label{eq:rfrf}
p \rightarrow \left\{ \begin{array}{ll} p-n \langle n,p \rangle + n
\langle n,p \rangle \sqrt{ 1 - 2 \frac{\Delta S}{\langle n,p \rangle ^2}} , & \textrm{if}\ \langle n, p
\rangle ^2 > 2 \Delta S \ \textrm{(refraction)} \\ p-2 n \langle n, p
\rangle, & \textrm{if}\ \langle n, p \rangle ^2 < 2 \Delta S \
\textrm{(reflection)} \end{array} \right. 
\end{equation}
Thus the trajectory will go through the topological sector boundary only if \(
\langle n, p \rangle^2 > 2 \Delta S \). In a HMC algorithm \( \langle n, p
\rangle = \mathcal{O} (1) \) and has exponential distribution.  \(\Delta S\),
however, is not the exact value of the height of the potential wall, but it is
the change of the pseudofermionic action at the boundary. From now on, we will
distinguish between these two quantities. We call the former \( \Delta
S_{\textrm{exact}}\) and the latter \(\Delta S_{\textrm{pf}}\).

Let us take a closer look\footnote{The following considerations in this
section have already partially appeared in \cite{DeGrand:2005vb}.}  on the
relation between \(\Delta S_{\textrm{exact}}\) and \(\langle \Delta
S_{\textrm{pf}} \rangle \).  In particular, we show that the jump in the
pseudofermionic action overestimates \(\Delta S_{\textrm{exact}} \).  Let us
assume that the trajectory crosses the boundary. Let \(H_{-}\) and \(H_{+}\)
be the overlap operator evaluated on the two sides of the boundary right
before and after the crossing, respectively. Clearly \(H_{-}\) and \(H_{+}\)
contain the same gauge configuration, but they differ, since one eigenvalue of
\(H_W\) changes sign on the boundary. In the HMC algorithm one chooses the
pseudofermion field as
\begin{equation*}
\phi=H_{-} \eta,\ \ \ \ \ \phi^\dagger=\eta ^\dagger H_{-}, 
\end{equation*}
where \(\eta, \eta ^\dagger\) are random vectors with Gaussian distribution,
in order to generate \(\phi, \phi^\dagger\) with the correct distribution. (In
a real simulation one chooses new pseudofermion configurations only at the
beginning of each trajectory, but for simplicity let's consider, that \(\phi\)
and \(\phi ^ \dagger\) are refreshed when hitting the boundary.)  The jump of
the pseudofermionic action now reads:
\begin{equation*}
\Delta S_{\textrm{pf}} =S_{\textrm{pf} +} - S_{\textrm{pf} -}= \eta^ \dagger (H_{-}
H^{-2}_{+} H_{-} -1) \eta
\end{equation*}
The relation between \(\Delta S_{\textrm{exact}}\) and \( \Delta
S_{\textrm{pf}}\) can be obtained by the following straightforward calculation:

\begin{equation*}
e^{- \Delta S_{\textrm{exact}}} = \frac{\det H^2_{+}}{\det H^2_{-}} =
\frac{\int [d \eta ^ \dagger] [d \eta] e^{- \eta^ \dagger \eta} e^{- \eta^ \dagger (H_{-} H^{-2}_{+} H_{-}
  -1) \eta}}{\int [d \eta ^ \dagger] [d \eta] e^{- \eta^ \dagger \eta}} =
\end{equation*}
\begin{equation*}
= \langle e^{-\eta^ \dagger (H_{-} H^{-2}_{+} H_{-} -1) \eta} \rangle 
_{\eta^{\dagger} \eta} \geq e^{- \langle \eta^ \dagger (H_{-} H^{-2}_{+} H_{-}
-1) \eta \rangle _{ \eta^{\dagger} \eta}} = e^{- \langle \Delta
S_{\textrm{pf}} \rangle } 
\end{equation*}
The inequality in the second line is a consequence of the concavity of the
\(e^{-x}\) function. So we conclude to:
\begin{equation*}
\langle \Delta S_{\textrm{pf}} \rangle \geq \Delta S_{\textrm{exact}}.
\end{equation*}

We can examine this relation in realistic simulations, if we take into
account, that there is a simple relation between \(H_{+}\) and
\(H_{-}\). Let's denote by \(\lambda_0\) the eigenvalue of \(H_W\) which
crosses zero at the boundary, and by \(|0 \rangle\) the eigenvector belonging
to \(\lambda_0\). With this notation:
\begin{equation*}
\label{eq:Hpm}
H_{+}=H_{-} + c |0 \rangle \langle 0 |,
\end{equation*}
where
\begin{equation*}
c=\Delta \textrm{sgn} \lambda_0 \ m_0 (1-\frac{m}{2 m_0}), 
\end{equation*}
with $\Delta \textrm{sgn} \lambda_0 = \pm 2$ being the jump of ${\rm sgn}
\lambda_0$ on the boundary.  The expectation value of the discontinuity in the
pseudofermionic action is:
\begin{equation*}
\langle \Delta S_{\textrm{pf}} \rangle = \langle \eta^ \dagger (H_{-}
  H^{-2}_{+} H_{-} -1) \eta \rangle _{ \eta^{\dagger} \eta} =
  \textrm{Tr}(H_{-} H^{-2}_{+} H_{-} -1) =
\end{equation*}
\begin{equation} 
\label{eq:Spf}
= \textrm{Tr} \big( (1-c |0 \rangle \langle 0 | H_{+}^{-1})(1-c\ H_{+}^{-1} |
0 \rangle \langle 0 | )-1 \big) = -2 c \langle 0 | H_{+}^{-1} | 0 \rangle + c^2
\langle 0 | H_{+}^{-2} | 0 \rangle. 
\end{equation}
In a similar way one can get a simple formula for the exact value of the jump
on the boundary:
\begin{equation}
\label{eq:Sex}
e^{-\Delta S_{\textrm{exact}}}=\frac{\det H_{+}^{2} }{\det H_{-}^{2}}=
\frac{1} {\det(H_{+}^{-1} H_{-})^2} = \frac{1}{\det(1- c H_{+}^{-1} |0 \rangle
\langle 0 |)^2} = \frac{1}{(1-c \langle 0 | H_{+}^{-1} | 0 \rangle)^2}.
\end{equation}
eq. (\ref{eq:Spf}) and eq. (\ref{eq:Sex}) offers a numerically fast way to
determine both action jumps, since one needs only one inversion of the overlap
operator to obtain both of them.

\FIGURE{
\label{fi:sc}
\includegraphics*[width=15.0cm,bb=18 433 591 718]{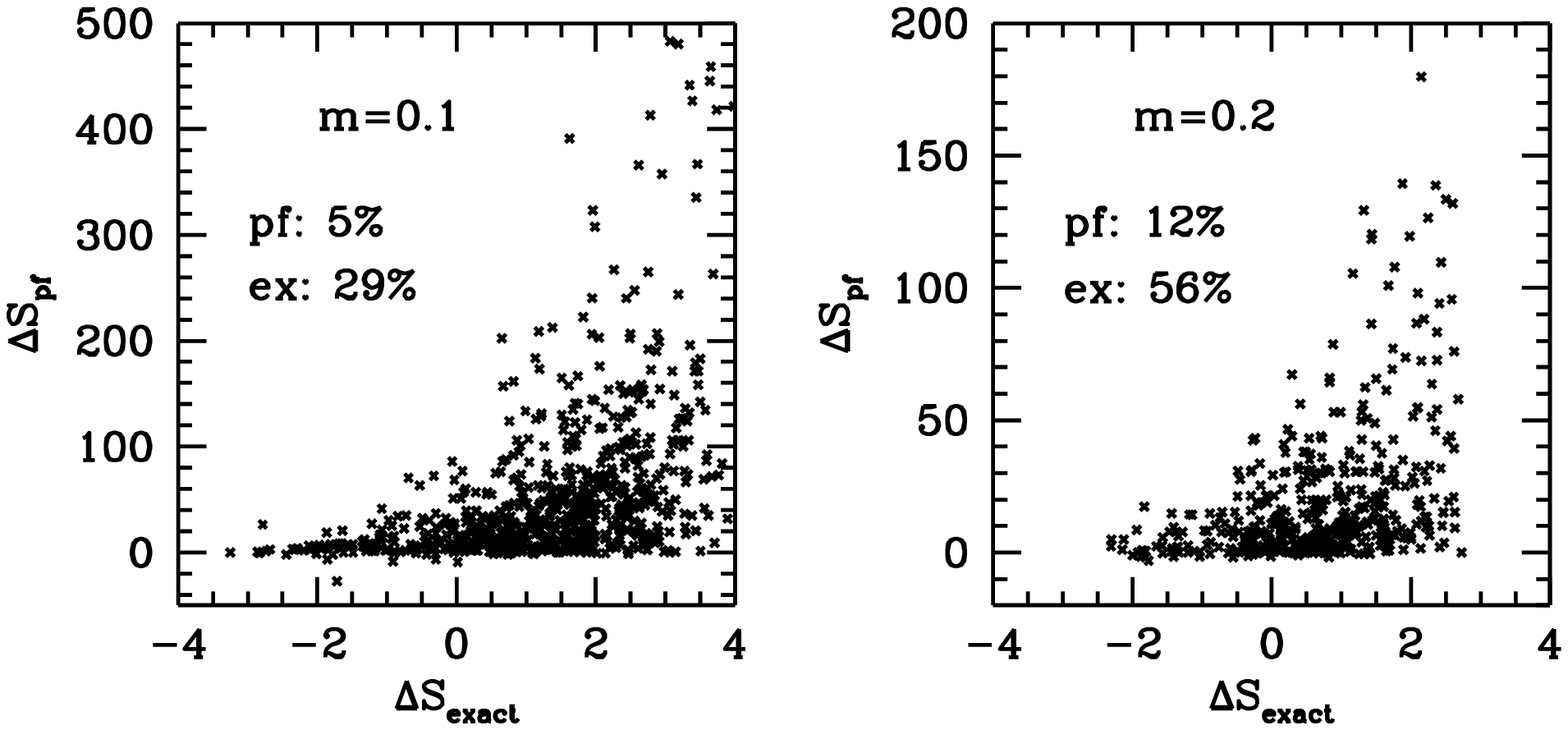}
\caption{The jump in the exact vs. pseudofermionic action at $\beta=4.05$ and
  $m=0.1, 0.2$. Since the average of $\langle n,p \rangle^2$ is around
  $\approx 1$, topological sector changing would happen considerably
  frequently using $S_{\rm exact}$, than with $S_{\rm pf}$. We also indicated
  the probability of topological sector changing with the pseudofermionic
  action, and an estimate on the probability using the exact action (assuming
  that the two algorithms would behave the same way except for the
  boundaries).  }  }

For illustration we made a scatter plot (Fig. \ref{fi:sc}) from a $6^4$
lattice at two different masses. (Details of our action will be described in
Section \ref{sec:num}.)  One can clearly see, that the use of the
pseudofermions has an awkward consequence: there are a huge amount of
crossings, where the topological sector changing fails only due to the
overestimation. One way to cure this is to use several pseudofermion
estimators instead of one \cite{DeGrand:2004nq}. More pseudofermions mean
smaller spread of the pseudofermionic action distribution, therefore the
overestimation is smaller, too.  However the computational time also increases
with the number of extra fields.  Obviously the best would be to use the exact
action in the simulations, but only its discontinuity on the boundary can be
calculated easily. The next section will present a technique, which uses this
discontinuity to get the relative weight of topological sectors.

\section{The new method}

\label{new}

In this section we propose a new method for the calculation of physical
observables by which it becomes possible to circumvent the problem of
topological sector changing described in the previous section. Let us write
the partition function in the form (assuming a vanishing \(\theta\)
parameter):

\begin{equation*}
Z=\sum_{Q=-\infty}^{\infty} Z_Q,
\end{equation*}
where $Z_Q$ is the partition function of the topological sector $Q$.
The expectation value of an observable:
\begin{equation*}
\langle O \rangle = \frac{\sum_{Q} Z_Q \langle O   \rangle_Q}{\sum_{Q} Z_Q}
= \frac{\sum_{Q} \frac{Z_Q}{Z_0} \langle O \rangle_Q}{\sum_{Q}
\frac{Z_Q}{Z_0}}, 
\end{equation*}
where the restricted expectation value \(\langle O \rangle_{Q}\) is
\begin{equation*}
\langle O \rangle_{Q}=\frac{1}{Z_Q}\int [dU]_Q O[U] \det H^2_Q \exp (-S_g).
\end{equation*}
For reasons which will be clear later the integration goes not only over the
configurations with $Q$ charge, but also over the boundary of the topological
sector as well (though the boundary has only zero measure in this case).  When
calculating the partition function in a given topological sector the following
boundary prescription is used: we define the determinant on the boundary as
the limit of determinants approaching the wall from the $Q$ side ($\det
H^2_Q$).  If the measurement of the quantities \(Z_{Q+1}/Z_{Q}\) would be
possible, then we could recover \(Z_{Q}/Z_{0}\) for any \(Q\). With these in
hand, we would need only the restricted expectation values \(\langle O
\rangle_{Q}\), whose measurement doesn't require topological sector changings.

Now we will show a way to measure \(Z_{Q+1}/Z_{Q}\).  It will make use of the
fact, that we can calculate easily \(\Delta S_{\textrm{exact}} \) on the
boundary of topological sectors (see eq. (\ref{eq:Sex})).  The pseudofermionic
action is only used to generate configurations in fixed topological sectors,
so its bad distribution for the jump of the action will not effect us.  (In
the following formulae $\Delta S$ will automatically mean $\Delta
S_{\textrm{exact}}$.)  The main idea is the following: an observable measured
in sector $Q$ is inversely proportional to $Z_Q$ and an observable in $Q+1$ is
to $Z_{Q+1}$. If the observables in the two sectors are concentrated only to
the common wall separating the two sectors, then from the ratio of the two
expectation values one can recover the ratio of the two sectors.

First let us measure in the $Q$ sector an operator, which is concentrated to
the boundary:
\begin{equation}
\label{eq:fd}
\langle \delta_{Q,Q+1} F\rangle_Q = \frac{1}{Z_Q}\int [dU]_{Q} \delta_{Q,Q+1}F[U]
\det H^2_Q
\exp (-S_g) 
, 
\end{equation}
where we introduced the distribution \(\delta_{Q,Q+1}\), a Dirac-\(\delta\),
which is equal to zero everywhere but on the \(Q, Q+1\) boundary.  Then let us
measure another operator $G$ on the same wall (thus on the boundary separating
sectors $Q$ and $Q+1$), but now from the $Q+1$ sector:
\begin{equation}
\label{eq:gd}
\langle \delta_{Q,Q+1} G\rangle_{Q+1} = \frac{1}{Z_{Q+1}}\int [dU]_{Q+1} \delta_{Q,Q+1}G[U]
\det H^2_{Q+1}
\exp (-S_g) 
. 
\end{equation}
The wall is the same (i.e. $[dU]_{Q}\delta_{Q,Q+1}=[dU]_{Q+1}\delta_{Q,Q+1}$)
in both cases, however due to our boundary prescription the determinants are
different on it. Therefore if $F$ and $G$ satisfies:
\begin{equation}
\label{eq:db1}
F[U]\det H^2_{Q}[U]=G[U]\det H^2_{Q+1}[U],
\end{equation}
then
the ratio of eq. (\ref{eq:fd}) and eq. (\ref{eq:gd})  gives us 
\begin{equation}
\label{eq:db2}
\frac{\langle \delta_{Q,Q+1} F\rangle_Q}{\langle \delta_{Q,Q+1}
G\rangle_{Q+1}}=\frac{Z_{Q+1}}{Z_Q}.
\end{equation}

The easiest choice is $G(U)=1$ and $F(U)=\det H^2_{Q+1}/\det H^2_Q = \exp
(-\Delta S)$, the ratio of sectors becomes:
\begin{equation}
\label{eq:zq}
Z_{Q+1}/Z_Q=\frac{\langle \delta_{Q,Q+1} \exp (-\Delta S) \rangle_Q}{\langle \delta_{Q,Q+1}\rangle_{Q+1}}.
\end{equation}
This choice is still not optimal, since the measurement of the numerator is
problematic, if the distribution of $\Delta S$ extends to negative values.
The exponential function amplifies the small fluctuations in the negative
$\Delta S$ region, which can destroy the whole measurement: a very small
fraction of the configurations will dominate the result.  As a consequence one
ends up with relatively large statistical uncertainties.  With a slightly
different choice of $F$ and $G$ we can improve on the situation. With
$F(U)=\Theta (\Delta S - x) \exp (-\Delta S)$ and $G(U)=\Theta (\Delta S - x)$
we can omit the problematic part of the $\Delta S$ distribution (the values
smaller than $x$) from the measurement, and we get:
\begin{equation}
\label{eq:zq1}
Z_{Q+1}/Z_Q=\frac{\langle \delta_{Q,Q+1} \exp (-\Delta S) \rangle_Q^{\Delta S >x}}{\langle
\delta_{Q,Q+1}\rangle_{Q+1}^{\Delta S >x}}.
\end{equation}  
The price of this choice of $F,G$ is that we do not make use of the $\Delta
S<x$ part of our data set.  The value of $x$ can be tuned to minimize the
statistical error.

Let us note that eq. (\ref{eq:db1}) can be viewed as a detailed balance
condition on a given $U$ configuration between $Q$ and $Q+1$ sector ($F$ and
$G$ are just the ``transition probabilities'').  This can give us a hint, that
the Metropolis-step is a good a solution for $F,G$: $F=\min(1,\exp(-\Delta
S))$ and $G=\min(1,\exp(\Delta S))$.  The ratio of sectors is simply:
\begin{equation}
\label{eq:zq2}
Z_{Q+1}/Z_Q=\frac{\langle \delta_{Q,Q+1} \min(1,\exp (-\Delta S)) \rangle_Q}{\langle \delta_{Q,Q+1} \min(1,\exp(\Delta S))\rangle_{Q+1}}.
\end{equation}
The inconvenient part of the distribution ($\Delta S<0$) is cut off, however
in contrast to eq. (\ref{eq:zq1}) all configurations are used to get the
expectation values.

We have achieved our main goal: without making expensive topological sector
changes we can obtain the ratio of sectors (see eq. (\ref{eq:zq},
\ref{eq:zq1}, \ref{eq:zq2})).  The key point is to make simulations
constrained to fixed topological charge, and match the results on the common
boundaries of the sectors.  In the next subsection we will show in the
framework of HMC, how to measure an expectation value, which contains a
Dirac-delta on the surface. Our proposal is that the generation of
trajectories inside a sector can be done using the pseudofermionic action, we
do not need there the exact action.  Since no sector changing is required, the
inconvenient distribution of the pseudofermionic action jump on the boundary
will not effect the measurement of the ratios of sectors.  The exact action is
needed only on the boundary: the formulas (\ref{eq:zq}, \ref{eq:zq1},
\ref{eq:zq2}) require \(\Delta S\).

It is clear that using the exact fermion action when measuring the ratio of
sectors outperforms the conventional HMC, where topological sector changing is
hindered by the distribution of the pseudofermionic action jump.  Even in that
(at the moment theoretical) case if we were able to use the exact fermionic
action in simulations, the above presented method is better in determining the
ratios of topological sectors. Consider a small quark mass simulation, where a
HMC using the exact fermionic action sticks into the trivial topological
sector (now due to the fact, that nontrivial topologies are suppressed by the
fermion determinant).  If the simulation time is not long enough, then we have
no information at all on the small (but nonvanishing) ratio $Z_1/Z_0$. However
in our method this small quantity can be measured.  The more we hit the wall
from the two sides the more precisely we can measure $Z_1/Z_0$.  The same
argument applies to an R-type algorithm (where it is possible to use the exact
$\Delta S$ jump, when crossing topological sectors \cite{DeGrand:2005vb}).

Obviously an important issue for this new method is whether topological
sectors defined by the overlap charge are path-connected or not.  We refer to
some results in the Abelian and in the non-Abelian gauge group case
\cite{Luscher:1998du,Adams:2002ms}.  If configurations with the same $Q$ would
not be continuously connectable in sector $Q$, then our assumption that we
make measurements on the common boundary of sectors could be violated. It
could happen, that the wall sampled from sector $Q$ does not coincide with the
wall sampled from $Q+1$.  Moreover the fixed sector simulations would also
violate ergodicity in this case. Let us note here that the large
autocorrelation time for the topological charge in the conventional
pseudofermionic HMC effectively also causes the breakdown of ergodicity. In
case of non-connected sectors one can cure these problems by releasing the
system from a sector after a certain amount of time and closing it to another.

\vspace{0.5cm}

\noindent \emph{Expectation value of a \(\delta\)-function}
\vspace{0.5cm}

In a HMC simulation one determines the expectation value of an operator by
calculating a sum over the measured values of the observable on $N$
configurations, which are generated with the proper weights, by which they
occur in the functional integral. In practice it is not possible to measure an
operator containing a Dirac-delta on the boundary surface on these
configurations, because none of them will be exactly located on it.  Therefore
we formulate a somewhat different measurement method, and discuss its
properties. As a result one has to make measurements at those points of the
trajectories, where they hit the boundary.

Let us see the details. Consider for a moment that we are able to integrate
the Hamiltonian equations of motion exactly.  If the distribution of the gauge
field, pseudofermion field and momenta was correct at the starting point of
the trajectory, then it will be still correct for any of the inner
points. This fact follows from energy and area conservation and
reversibility. So we make no mistake if we put the inner points of the
trajectory into the ensemble.  We can write:
\begin{equation}
\langle O \rangle= \lim_{N\to \infty}\frac{1}{N} \sum_{i=1}^{N} O_i=\lim_{N\to \infty} \frac{1}{N} \sum_{i=1}^{N}
\int_i  O(U_t) dt=\lim_{N \to \infty}\frac{1}{N} \sum_{i=1}^{N} O_{i,i+1}
\end{equation} 
where \(t\) is the microcanonical time, \(U_t\) is the gauge configuration at
time \(t\), the \(i\) subscript at the integral means that the integration
goes for the \(i\)th trajectory, and \(O_{i,i+1}\) is thus the average of the
operator \(O\) along this trajectory.  In the case of an observable, which
contains a \(\delta\)-function, like those in eq. (\ref{eq:zq1}) or in
eq. (\ref{eq:zq2}) this reads:
\begin{equation}
O_{i,i+1} = \int_i dt\ o(U_t)\ \delta \left(\lambda_0(U_t)\right),
\end{equation}
where it is indicated, that the \(\delta\)-function depends on the gauge
configuration only through the smallest absolute value eigenvalue
\(\lambda_0\) of \(H_W\). If the variable of integration is changed from \(t\)
to \(\lambda_0\), we get:
\begin{equation}
O_{i,i+1} = \int d\lambda_0 \sum_j \left| \frac{dt}{d\lambda_0} \right|_{t_j,
  \lambda_0(U_{t_j})=0} o(U_{t_j}) \delta(\lambda_0).
\end{equation}
We can go further by determining the time derivative of the smallest
eigenvalue:
\begin{equation}
\frac{d\lambda_0}{dt}= \langle 0 | \textrm{tr} \frac{\partial
H_W}{\partial U^T}
\frac{d U}{d t} |0 \rangle
\end{equation}
where, again \(|0 \rangle\) is the eigenvector belonging to \(\lambda_0\). The
trace and transpose operations are meant to be in color space. Recognizing,
that with our previous notation
\begin{equation*}
\frac{d U}{d t}=p U \ \ \ \textrm{and} \ \ \ \langle 0 | U
\frac{\partial H_W}{\partial U^T} | 0 \rangle = n
\end{equation*}
yields
\begin{equation}
\frac{d\lambda_0}{dt}=\langle n,p \rangle, \ \ \ \left| \frac{dt}{d \lambda_0}
\right| =\frac{1}{|\langle n,p \rangle|},
\end{equation}
\begin{equation}
\langle O \rangle =\lim_{N\to \infty} \frac{1}{N} \sum_{i=1}^{N} \sum_j \left. o(U_{t_j}) \frac{1}{| \langle n,p_{t_j}
  \rangle |} \right| _{t_j, \lambda_0(U_{t_j})=0}
\end{equation}
If we put it simple, the above formula says, that since the integration is in
microcanonical time, the angle and velocity by which the trajectory hits the
boundary has to be taken into account.

Let us turn back to the case, when the integration of the equations of motion
can be done only with finite step size integrator. The leap-frog procedure has
$O(\epsilon ^3)$ error per microcanonical step, which after $1/\epsilon$ steps
makes the trajectory differ by $O(\epsilon ^2)$ from the exact trajectory. If
we can guarantee, that the modification of the trajectory at the boundary also
violates the equations of motions only upto $O(\epsilon ^2)$, then in the
final results the errors will be proportional with $O(\sqrt{N}\epsilon
^2)$. The original reflection algorithm in \cite{Fodor:2003bh} has
$O(\epsilon)$ errors, later it was improved to $O(\epsilon ^2)$ in
\cite{Cundy:2005pi}.  In the Appendix we propose a different reflection
procedure and prove that it is reversible, area-preserving and conserves the
energy upto $O(\epsilon ^2)$.

\section{Numerical results}
\label{sec:num}
In the previous section we described a new method, to solve 
the topological sector changing problem of pseudofermionic HMC simulation.
We describe here the details of the simulations, and finally
give the topological susceptibility in physical units measured on $8^4$ and
$8^3\times 16$ lattices. 

Simulations were done using unit length trajectories, separated by momentum
and pseudofermion refreshments. The system was confined to a fixed topological
sector in each run, we reflected the trajectories whenever they reached a
sector boundary.  The end points of the trajectories obviously follow the exact
distribution in a given sector, usual quantities can be measured on them.
When calculating the ratio of sectors using eq. (\ref{eq:zq}) or
eq. (\ref{eq:zq1}) or eq. (\ref{eq:zq2}) we integrated along the trajectories,
this quantity will be burdened by a step size error.

\FIGURE{
\includegraphics*[width=15.0cm,bb=18 433 591 718]{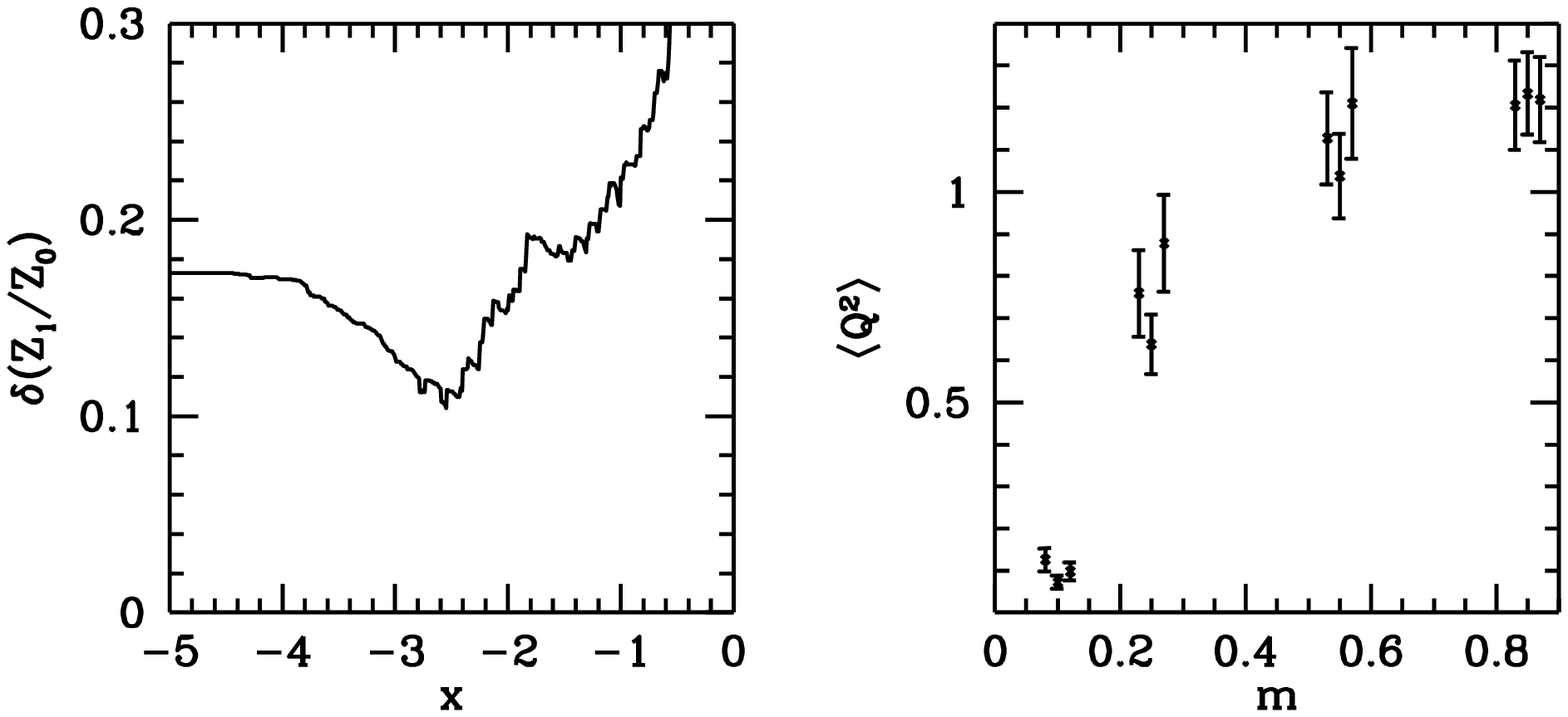}
\caption{
\label{fi:test}
{\it Left panel:} a typical optimization procedure of the lower limit ($x$) on
$\Delta S$ in the formula (\ref{eq:zq1}).  The statistical error of the ratio
$Z_1/Z_0$ shows a minimum as the function of $x$, which is considered as the
optimal value. {\it Right panel:} Bare mass dependence of topological
susceptibility using three different methods on $6^4$ lattices.  The points
corresponding to the same mass were slightly shifted vertically for
clarity. Result based on our new technique and eq. (\ref{eq:zq1}) is on the
left, based on eq. (\ref{eq:zq2}) is in the middle, the standard
pseudofermionic HMC is on the right.  The simulation parameters are from
\cite{Fodor:2004wx} } }

In case of large enough statistics the value of $Z_{Q+1}/Z_Q$ should be the
same, independently which of the three formula (\ref{eq:zq}, \ref{eq:zq1},
\ref{eq:zq2}) was used to calculate it.  We omit eq. (\ref{eq:zq}) in the
following, since it is hard to give a reliable error estimate on the
expectation value of $\exp (-\Delta S)$, if $\Delta S$ can be arbitrary
negative number. Eq. (\ref{eq:zq1}) still measures $\exp (-\Delta S)$, but
with a lower limit ($x$) on $\Delta S$. Smaller limit yields a smaller and
more reliable error, however the statistics is decreased at the same time.
One can tune the value of $x$, so that the statistical error takes its
minimum. A result of a typical optimum search can be seen on the left panel of
Fig. \ref{fi:test}.  The optimal value can be compared to the one obtained
from eq. (\ref{eq:zq2}).  On the right panel of Fig.  \ref{fi:test} the two
new topological susceptibilities and the one calculated by using traditional
pseudofermionic HMC \cite{Fodor:2004wx} are shown. The agreement is perfect.

\TABLE{
\begin{tabular}{|c|c|c|c|}
\hline
m    & $r_0$ & $m_\pi$ &  $L m_\pi$    \\
\hline
$0.03$ &   $3.52(13)$  & $0.29(11)$  &  $2.4$   \\
\hline
$0.1$  &   $3.17(5)$  & $0.53(4)$ & $4.3$ \\
\hline
$0.2$  &  $2.89(2)$ & $0.74(6)$    & $5.9$ \\
\hline
$0.3$  & $2.88(6)$ & $0.99(8)$    & $7.9$ \\
\hline
\end{tabular}
\caption{
\label{tb:spect} Sommer-scale, pion mass and pion mass times box size on $8^3\times 16$ lattices. 
}
}

To measure the topological susceptibility on $8^4$ lattices we generated
configurations with tree-level Symanzik improved gauge action ($\beta=4.15$
gauge coupling) and 2 step stout smeared overlap kernel ($\rho=0.15$ smearing
parameter, the kernel was the standard Wilson matrix with $m_0=1.3$). We
performed runs in sectors $Q=0\dots 3$ (based on the measured $Z_3/Z_2$ we can
conclude, that the contribution of $Q \ge 4$ sectors are small compared to
statistical uncertainties).  For the negatively charged sectors we used the
$Q\to -Q$ symmetry of the partition function. The bare masses were $m=0.03,
0.1, 0.2$ and $0.3$, at each mass approximately 1000 trajectories were
collected. The average number of the topological sector
boundary hittings was
around $1.5$ per trajectory. We calculated the ratio of sectors using
eq. (\ref{eq:zq1}) and eq.  (\ref{eq:zq2}). The result for the topological
susceptibility can be seen on Fig. \ref{fi:ts}.  It is nicely suppressed for
the smallest mass.  To convert it into physical units, we measured the static
potential and the pion mass on $8^3 \times 16$ lattices (Table
\ref{tb:spect}). Since our statistics was quite small on these asymmetric
lattices, the errors are large.  Note, that in order to get the mass-dimension
4 topological susceptibility in physical units, one has to make very precise
scale measurements.  

\FIGURE{ \includegraphics*[width=15.0cm,bb=18 429 591
717]{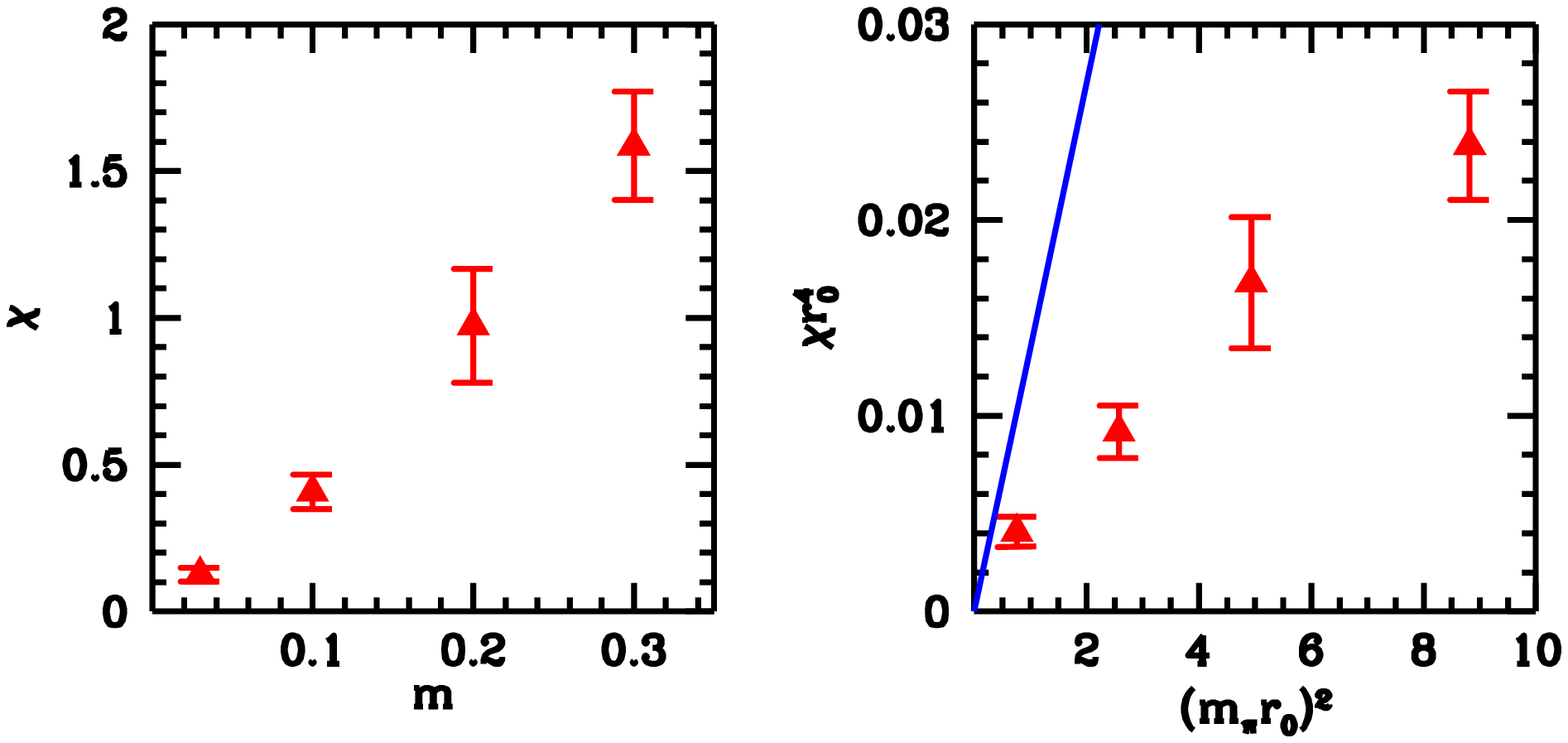}
\caption{
\label{fi:ts}
Topological susceptibility as the function of quark mass on $8^4$ lattices in
lattice units (left), and in physical units as the function of pion mass
(right).  Scale fixing and mass measurements were done on $8^3\times 16$.  The
error bars on the right plot do not contain the errors of scale fixing.  The
line is the leading order chiral behavior in the continuum.  }  } When
interpreting the results, one should keep in mind, that the volume is small,
and the lattice spacing is large. Note however, that smeared kernel overlap
actions show nice scaling behavior and good locality properties already at
moderate lattice spacings\cite{Kovacs:2002nz,Durr:2005an}.
 
\label{num}

\section{Conclusions}
\label{conclusions}
In this paper we studied the problem of topological sector changing in
dynamical overlap simulations.  The origin of the unexpectedly large
autocorrelation time for the topological charge is connected to
pseudofermions, which approximate the fermion determinant. The pseudofermionic
action overestimates the size of the discontinuity in the fermion determinant
at the topological sector boundary, so the system cannot enter easily to a new
topological sector. This happens even if the use of the exact determinant
favored a transition. (The discontinuity of the exact determinant can be
calculated in a rather inexpensive way.)

We developed a new method, which circumvents the problem of topological sector
changing. It confines the system to fixed topological sectors (by always
reflecting the HMC trajectories from the topological sector boundaries). Thus
overestimating the discontinuity of the determinant due to pseudofermions will
not effect the determination of topology related quantities. The relative
weight of two topological sector is obtained by measuring appropriate
operators on the common boundary surface. The measurement of such operators
can be carried out by extending the usual HMC measurement method, however an
$O(\epsilon^2)$ extrapolation in HMC step size is required.

The new method was tested on $6^4$ lattices, where previous conventional HMC
results were available. The old and new results were consistent. We also
measured the topological susceptibility on $8^4$ lattices with an improved
overlap fermion and gauge action, furthermore simulations were done on
$8^3\times 16$ lattices to convert the lattice results into physical units.

\vspace{0.5cm}  
\noindent
{\bf Acknowledgments:\\} 
Useful comments on the manuscript from A. D. Kennedy and
discussions with T. G. Kov\'acs
are acknowledged. For this work a modified version of the MILC
Collaboration's public code \cite{MilcCode} with SSE instructions
\cite{Fodor:2002zi} was used.  Simulations were carried out on the ALICENext
PC-Cluster (1024 AMD-Opteron processors) at Wuppertal University, Germany and
on the PC-cluster (330 Intel-P4 nodes) at the E\"otv\"os University of
Budapest, Hungary.  This work was partially supported by Hungarian Scientific
grants, OTKA-T34980/\-T37615/\-M37071/\-T032501/\-AT049652.
This research is part of the EU Integrated Infrastructure
Initiative Hadronphyisics project under contract number
RII3-CT-20040506078.

\section*{Appendix}
We present a reflection algorithm which conserves the energy upto
$O(\epsilon^2)$, and which is reversible and area conserving. Until the
boundary the trajectory is evolved by the usual leapfrog procedure. An
elementary leapfrog step can be written in the following symbolic way:
\begin{equation}
P(\epsilon/2)U(\epsilon) P(\epsilon/2),
\end{equation} 
where $U(\epsilon)$ and $P(\epsilon)$ are the operators updating the links
($u$) and the momenta ($p$):
\begin{align*}
U(\epsilon): u \to \exp (\epsilon p) u, &&
P(\epsilon): p\to p - \epsilon A \left[u\frac{\partial}{\partial u} \left(S_g +
S_{\rm pf}\right)\right],
\end{align*}
where in the force term the $A$ operator projects onto traceless,
antihermitian matrices (in color indices).  Now we split the evolution of the
links into two parts:
\begin{equation}
\label{eq:lf}
P(\epsilon/2)U(\epsilon/2) \cdot U(\epsilon /2)P(\epsilon/2).
\end{equation} 
Consider that the boundary would be crossed during one of the evolutions of
the links in eq. (\ref{eq:lf}). Then replace the original leapfrog with the
following:
\begin{equation*}
P(\epsilon/2)U(\epsilon/2) \cdot U(\epsilon_c) P(\epsilon_c) R P(\epsilon_c) U(\epsilon_c) \cdot
U(\epsilon/2) P(\epsilon/2),
\end{equation*}
where $R$ is simply the reflection of the momenta (eq.  (\ref{eq:rfrf})).
$\epsilon_c$ is the time to reach the boundary surface measured from the
midpoint of the leapfrog. Thus if the crossing would happen in the first
evolution then $\epsilon_c<0$, if in the second, then $\epsilon_c>0$.  The
reversibility can be checked easily, for the energy conservation we note that
both $U(\epsilon)P(\epsilon)$ and $P(\epsilon)U(\epsilon)$ conserves the
energy upto $O(\epsilon^2)$. If the evolution time does not depend on the
variables, then both $U(\epsilon)$ and $P(\epsilon)$ are area conserving. The
only problem arises due to the link and momenta dependence of $\epsilon_c$.
We will now show that the combined effect of the $\epsilon_c$ updates
\begin{equation}
\label{eq:corr}
U(\epsilon_c)P(\epsilon_c) R P(\epsilon_c)U(\epsilon_c)
\end{equation}
is still area preserving.

For simplicity we will not carry out here the computation using the $SU(3)$
structure, but only for a $k$-dimensional Euclidean vectorspace (the $q$
coordinates are $k$-component vectors). All features of the proof are also
present in this simpler case. We have to calculate the Jacobian of the
following transformation
\begin{align*}
q'(q,p)=q+\epsilon_c p+\epsilon_c p' &&
p'(q,p)=Rp+h.
\end{align*}
$R_{ab}=\delta_{ab}-2 n_a n_b$ is the reflection operator ($a,b=1\dots k$),
where $n_a$ is the normal vector of the surface.  $h$ can be expressed as
$h=-\epsilon_c F - \epsilon_c RF$, where $F$ is the force on the wall
(interpreted in the sector into which we reflect back the trajectory). Since
$h$ is measured on the wall, its $q$ and $p$ derivatives satisfy $\partial h/
\partial p=\epsilon_c\partial h/ \partial q$ (see
\cite{Fodor:2003bh}). Furthermore $h$ is orthogonal to $n$ ($(nh)=0$).  For
the $q$ derivative of $p'$ we introduce the matrix $Z_{ab}$, which contains
the derivatives of the normal vector and $h$:
\begin{align*}
\der{p'_a}{q_b}=Z_{ab}.
\end{align*}
Using that $n$ and $h$ are functions of the wall coordinates, $\epsilon_c Z$
will appear in the $p$ derivative of $p'$:
\begin{align*}
\der{p'_a}{p_b}=R_{ab}+\epsilon_c Z_{ab}.
\end{align*}
The partial derivatives of $q'$ are:
\begin{align*}
\der{q'_a}{q_b}=\delta_{ab}+(p_a+p'_a)\der{\epsilon_c}{q_b}+\epsilon_c\der{p'_a}{q_b},\\
\der{q'_a}{p_b}=\epsilon_c\delta_{ab}+(p_a+p'_a)\der{\epsilon_c}{p_b}+\epsilon_c\der{p'_a}{p_b}.
\end{align*}
With the help of the following identities (see \cite{Fodor:2003bh}):
\begin{align*}
\der{\epsilon_c}{q_a}=-\frac{n_a}{(np)} && \der{\epsilon_c}{p_a}=-\epsilon_c
\frac{n_a}{(np)},
\end{align*}
the Jacobian can be written in a hypermatrix form (${x \circ y}$ has
components $x_ay_b$):
\begin{align*}
J=
\begin{pmatrix}
\der{q'}{q} & \der{q'}{p}\\
\der{p'}{q} & \der{p'}{p}\\
\end{pmatrix}=
\begin{pmatrix}
{ 1} & -2\epsilon_c { Q}\\
{ 0} & { R} 
\end{pmatrix}-
\frac{1}{(np)}
\begin{pmatrix}
1 & \epsilon_c\\
0 & 0
\end{pmatrix}\otimes
[ (p+p') \circ n]+
\begin{pmatrix}
\epsilon_c & \epsilon_c^2 \\
1 & \epsilon_c
\end{pmatrix}
\otimes { Z}.
\end{align*}
$Q$ is the projector to the subspace orthogonal to $n$ ($Q=1-n\circ n$).  $J$
can be written as a product of two matrices $J=J_1J_2$, where
\begin{align*}
J_1=\begin{pmatrix}
{ 1} & -2\epsilon_c { Q}\\
{ 0} & { R}
\end{pmatrix}, 
\end{align*}
having determinant $-1$ (since $R$ is a reflection).
\begin{align*}
J_2=\begin{pmatrix}
1 & 0\\
0 & 1
\end{pmatrix}\otimes { 1}
-\frac{1}{(np)}
\begin{pmatrix}
1 & \epsilon_c \\
0 & 0 		
\end{pmatrix}\otimes [(p+p') \circ n]
+
\begin{pmatrix}
-\epsilon_c & -\epsilon_c^2 \\
1 & \epsilon_c
\end{pmatrix}
\otimes { RZ},
\end{align*} 
which has an upper triangular form in the $2\times 2$ space in a suitable
basis. The determinant of $J_2$ will be unity due to the orthogonality of
$p+p'=p+Rp+h=-2Qp+h$ and $n$.  Therefore $\det J=-1$, which shows that the
transformation (\ref{eq:corr}) preserves the area.
\end{document}